\documentclass[12pt]{article}
 \usepackage[dvips]{graphicx}
\begin{document}
 \setcounter{section}{0}
\title{On a loss of information in a transition from quantum to a quasi-classical regime}
\author{ Alex Granik\thanks{Department of Physics,UOP,Stockton,CA.95211;E-mail:agranik@uop.edu}}
\date{}
 \maketitle
\begin{abstract}
By defining information entropy in terms of probabilities
densities $|\Psi|^2$ ($\Psi$ is a wave function in the coordinate
representation) it is explicitly shown how a loss of quantum
information occurs in a transition from a quantum to a
quasi-classical regime.
\end{abstract}
 The Boltzmann-Gibbs-Shannon entropy
\begin{equation}
\label{eq:1} I=-\sum_{i=1}^Np_iLogp_i
\end{equation}
($p_i$ is the classical probability) has an obvious relation to
von Neumann entropy $$S=-Tr\rho Log\rho$$ where the diagonal terms
of the density matrix $\rho$ are the probabilties $p_i$.\

Usually in quantum mechanics the probability $p_i$ is considered
as a function of energy $E_i$ (e.g. \cite{LL}). Since we study a
rather narrow problem of information dynamics  when the
probability is described in terms of the wave function $\Psi$, we
will not discuss a more sophisticated (and complete) approach
based on the density matrix and information processing involving
quantum bits (qubits) ( e.g., \cite{CA}).

To achieve our goal ,we proceed along the lines used in derivation
of continuity equation for probability density. To this end
consider the probability in its coordinate representation. Thus
$p_i$ can be viewed as a probability of finding a particle in a
spatial interval $\Delta q_i=q_{i+1}-q_i$:
\begin{equation}
\label{eq:2}
 p_i=\int_{q_i}^{q_{i+1}} \Psi(q)^*\Psi(q) dq
\end{equation}
Using the mean value theorem, we rewrite Eq.(\ref{eq:2}) as
follows:
\begin{equation}
\label{eq:3} p_i=|\Psi_i(\tilde{q}_i)|^2\Delta q_i,~~q_i\leq
\tilde{q_i}\leq q_{i+1}
\end{equation}
where the sum of $p_i$ over all spatial intervals is
\begin{equation}
\label{eq:4} \sum_{i=1}^{\infty} p_i
=\sum_{i=1}^{\infty}|\Psi_i(\tilde{q}_i)|^2\Delta q_i=1
\end{equation}\\

Using Eq.(\ref{eq:3}) we get
\begin{equation}
\label{eq:5} p_iLog p_i=|\Psi_i|^2\Delta q_i[Log(|\Psi_i|^2)+Log
(\Delta q_i)]
\end{equation}
Now we take the following limit of (\ref{eq:5}) $$\Delta q_i
\rightarrow 0; ~~~\Delta q_i \rightarrow dq$$ As a result,
Eq.(\ref{eq:5}) yields
\begin{equation}
\label{eq:6} p_iLog
p_i\rightarrow|\Psi(q)|^2dq(Log|\Psi(q)|^2-1)+O(|\Delta q|^2)
\end{equation}

and equation (\ref{eq:1}) becomes respectively:
\begin{equation}
\label{eq:7} I=-\int|\Psi(q)|^2(Log|\Psi(q)|^2-1)dq
\end{equation}
Therefore we narrow our discussion even more by restricting it
with continuous spectra.

In a 3-D case the integration Eq.(\ref{eq:7}) is carried over a
spatial volume $d^3q.$ This means that the expression
\begin{equation}
\label{eq:8} \rho_I= -|\Psi(q)|^2(Log|\Psi(q)|^2-1)
 \end{equation}
 can be interpreted as the volume information density
 \begin{equation}
 \label{eq:A1}
 \rho_I=dI/dV
 \end{equation}

 We find its time variation:
 \begin{equation}
 \label{eq:9}
 \frac{\partial \rho_I}{\partial t}=
 -\frac{\partial|\Psi(q)|^2(Log|\Psi(q)|^2-1)}{\partial
 t}=-\frac{\partial |\Psi|^2}{\partial t}Log|\Psi|^2
 \end{equation}
 If we use the continuity equation for probability density
 (e.g.,\cite{LL1})
 $$\frac{\partial|\Psi|^2}{\partial t}=-
 \frac{\hbar}{2im}div(|\Psi|^2\nabla Log\frac{\Psi}{\Psi^*})$$
 then we obtain from Eq.(\ref{eq:9})
\begin{eqnarray}
\label{eq:10}
\frac{\partial \rho_I}{\partial t}=
 \frac{\hbar}{2im}div(|\Psi|^2\nabla Log\frac{\Psi}{\Psi^*})Log|\Psi|^2 \equiv\nonumber \\
 div(|\Psi|^2Log|\Psi|^2\frac{\hbar}{2im}\nabla
 Log\frac{\Psi}{\Psi^*})-\frac{\hbar}{2im}\nabla
 Log\frac{\Psi}{\Psi^*}\bullet \nabla |\Psi|^2
\end{eqnarray}
Inserting the definition of the information density $\rho_I$,
Eq.(\ref{eq:8}), and denoting the probability density
$|\Psi|^2=\rho$ into the first term on the right hand side of
Eq.(\ref{eq:10}) , we obtain
\begin{equation}
\label{eq:11} \frac{\partial\rho_I}{\partial t}
=-div[\frac{\hbar}{2im}(\rho_I-\rho)\nabla
Log\frac{\Psi}{\Psi^*}]-\frac{\hbar}{2im}\nabla
 Log\frac{\Psi}{\Psi^*}\bullet \nabla |\Psi|^2
 \end{equation}\\

 Let us consider the transition to the classical case by
 representing the wave function as follows
\begin{equation}
\label{eq:12} \Psi=\sqrt\rho e^{iS/\hbar}
\end{equation}
Using (\ref{eq:12}) in (\ref{eq:11}), we get
\begin{equation}
\label{eq:13} \frac{\partial\rho_I}{\partial t} =-div[\frac{\nabla
S}{m}(\rho_I-\rho)]-\frac{\nabla S}{m}\bullet \nabla \rho
\end{equation}
In the classical limit $\hbar t_c/m L_c^2 \rightarrow 0$ ( where
$L_c$ is a characteristic length and $t_c$ is the characteristic
time) phase $S$ becomes a classical action. As a result $\nabla
S/m =\vec{v}$, that is a classical particle velocity. This means
that in this limit Eq.(\ref{eq:13}) yields
\begin{equation}
\label{eq:14} \frac{\partial\rho_I}{\partial t}
=-div[\vec{v}(\rho_I-\rho)]-\vec{v}\bullet \nabla\rho
\end{equation}\\

If we integrate (\ref{eq:14}) over a volume $V$, use definition of
the information entropy (\ref{eq:A1}), and Gauss's theorem, we
obtain
\begin{equation}
\label{eq:15} \frac{\partial}{\partial t}\int\rho_i
dV=\frac{\partial I}{\partial t}= -\oint
(\rho_I-\rho)\vec{v}\bullet \vec{dA}- \int \vec{v}\bullet
\nabla\rho dV
\end{equation}

By taking the boundary of the volume $V$ to infinity and assuming
that $\rho(q\rightarrow\infty)\rightarrow 0$ we obtain from
(\ref{eq:15})
\begin{equation}
\label{eq:16} \frac{\partial I}{\partial t}= -\int \vec{v}\bullet
\nabla\rho  dV
\end{equation}

This means that in a transition from  a quantum to classical
regime the information entropy $I$ [defined in the above narrow
sense, Eq.(\ref{eq:7})] is not conserved. Instead, if the
classical velocity is in the direction of an increase of
probability density, the information entropy decreases.
Inversely,if the classical velocity is in the direction of a
decreasing probability density, the information density
increases.\\

These results are in agreement with the meaning of the information
entropy as a number of states accessible to a system. In a
transition to a classical regime this number drastically
decreases, thus signaling a decrease in information entropy. Quite
in agreement with that, and from another point of view, the less
probable states carry more information than the more probable
states.

One can interpret these results as a  statement that a classical
regime has less degree of freedom than its quantum counterpart.
This is definitely true when there is such a counterpart.\

As we have already stated earlier, the above definition of
information entropy is not fully appropriate for quantum mechanics
because it does not take into account the information carried by a
quantum phase. This serves as a strong indication that in the
quantum region one needs to use another, more general, definition
of information entropy , not necessarily associated with the
qubits, which will account for the information associated with the
quantum phase \cite {AG}.\\

{\bf{Acknowledgments}}

The author thanks V.Panico and C.Wulfman for the illuminating
discussions of the results.\\

\end{document}